# Competing SDW Phases and Quantum Oscillations in (TMTSF)$_2$ClO$_4$ in Magnetic Field

Danko Radić[1], Aleksa Bjeliš[1] and Dražen Zanchi[2]

[1] *Department of Physics, Faculty of Science, University of Zagreb, Zagreb, Croatia*
[2] *Laboratoire de Physique Théoretique et Hautes Energies, Universités Paris VI Pierre et Marie Curie – Paris VII Denis Diderot, Paris, France*

**Abstract:** We propose a new approach for studying spin density waves (SDW) in the Bechgaard salt (TMTSF)$_2$ClO$_4$ where lattice is dimerized in transverse direction due to anion ordering. The SDW response is calculated in the matrix formulation that rigorously treats the hybridization of inter-band and intra-band SDW correlations. Since the dimerization gap is large, of the order of transverse bandwidth, we also develop an exact treatment of magnetic breakdown in the external magnetic field. The obtained results agree with the experimental data on the fast magneto-resistance oscillations. Experimentally found 260T rapid oscillations and the characteristic $T_c$ dependance on magnetic field of relaxed material are fitted with our results for anion potential of the order of interchain hopping.

Numerous low temperature measurements on (TMTSF)$_2$ClO$_4$ in the last decade showed that the transverse lattice dimerization due to the anion ordering generates a FISDW ordering that is qualitatively different from that observed in other Bechgaard salts and explained within the well-known standard model of imperfect nesting [1]. Theoretical works [4, 5, 6] on this problem started from the extended version of the standard model with two pairs of quasi one-dimensional Fermi surfaces separated by the dimerization gap *V*, and analyzed the ensuing competition of intra-band (SDW$_\pm$) and inter-band (SDW$_0$) field induced orderings, the latter being the standard SDW in the absence of dimerization. However, these works, as well as some other interpretations [2], did not take into account that all these SDWs are the result of *hybridization* of inter-band and intra-band collective electron-hole excitations that cannot be neglected irrespectively to the value of *V*. The rigorous treatment of this hybridization goes through the matrix form of SDW susceptibility, developed recently [7, 8] at the RPA level. Furthermore, as follows from recent data [3], *V* is large, of the order of transverse bandwidth $t_b$. Since then one cannot use the usual perturbative inclusion of tunneling between close Fermi sheets [5], we introduce instead an exact treatment of the magnetic breakdown through the dimerization gap [9], achieving so a complete non-perturbative treatment of all effects of dimerizing gap within the extended standard model.

The phase diagram in the absence of magnetic field *B* is shown in Fig.1a. SDW$_0$ ordering is anti-ferromagnetic in the transverse dimerizing direction, i.e. the spin orientations on neighboring chains are opposite. It is stable in the range of small values of the ratio $V/t_b$. This range further narrows with the increase of $t'_b$, the imperfect nesting parameter due to the effective hopping between next nearest chains. On the other hand SDW$_\pm$ ordering is stable for large values of *V*, and is not affected by $t'_b$ at all. SDW$_\pm$ are superpositions of two anti-ferromagnetic orderings with generally different amplitudes, situated at nonequivalent sublattices in the dimerized crystal. A more precise value of the corresponding wave vector of ordering depends on the fine details of the plateau of effective imperfect nesting due to a finite dimerizing gap *V*. For intermediate values of *V* (0.1 < $V/t_b$ <1.6) there is a "valley" between SDW$_0$ and SDW$_\pm$ where metallic phase remains stable down to *T*=0. This "theoretical" range of possible values of *V* in slowly relaxed samples of (TMTSF)$_2$ClO$_4$ with no SDW ordering at *B*=0 is in agreement with direct experimental evidence into the value of *V*.

Passing to the regime of intermediate and strong external magnetic field in which experiments show FISDW ordering, we first consider the exact quantum mechanical solution of magnetic breakdown problem for electron tunneling through the dimerization gap in which two characteristic quantities appear to be relevant. The first is magnetic breakdown parameter $\kappa \equiv 2\omega_c t_b/V^2$ that determines electron quasi-classical over-gap tunneling probability $P=\exp(-\pi/2\kappa)$, $\omega_c$ being the cyclotron frequency. [5] The second quantity is dimensionless variable $r \equiv \sqrt{(0.77V)^2 + t_b^2}/\omega_c$. Rapid oscillations in transport properties are related to oscillations in electronic spectrum [5]. Indeed, we find that the electronic energy tilt due to anion ordering is periodic function in $r$, with period of about 0.8 [9]. Putting $v_F = 2 \cdot 10^5$m/s, $t_b$=300K this periodicity fits 260Tesla rapid oscillations for $V \approx 0.8 t_b$ and magnetic breakdown parameter $\kappa \approx 1$ for $B \approx 60$T.

The numerical insight into the diagonalized SDW matrix susceptibility shows the variety of possibilities as $V$ increases along the valley in Fig.1a. For values of V close to the left side of this range one has the SDW$_0$ ordering with cascading behaviour of $T_c(B)$ associated with $N$=1,3,5... quantum numbers (Fig. 1b). As $V$ increases SDW$_0$ and SDW$_\pm$ start to compete, resulting to the appearance of the 1st order phase transition from SDW$_0$ to SDW$_\pm$ at some $V$-dependent critical field $B_c$ (arrows in Fig. 1c). At higher values of $B$ SDW$_\pm$ attains a plateau of almost constant $T_c(B)$, whose height depends on $V$. For even higher values of $V$, those close to the right side of the valley in Fig. 1a, SDW$_0$ is completely suppressed and whole phase diagram is covered by SDW$_\pm$. The details of phase diagrams in each of above regimes are included into a more complete discussion [9], which shows a qualitative agreement with recent experimental results [4] on the evolution of the phase diagram with cooling speed, i. e. as the strength of anion ordering presumably decrease. We also note that the degeneracy of SDW$_+$ and SDW$_-$ phases can be straightforwardly lifted by taking into account non-linearity of longitudinal band dispersion. After the splitting of SDW$_+$ and SDW$_-$ the dominant instability will be SDW$_-$ while the SDW$_+$ sub-phases are expected to appear at lower temperatures, each nesting its own pair of Fermi sheets. Such scenario is impossible for SDW$_0$ since it nests all four Fermi sheets at a *single* critical temperature.

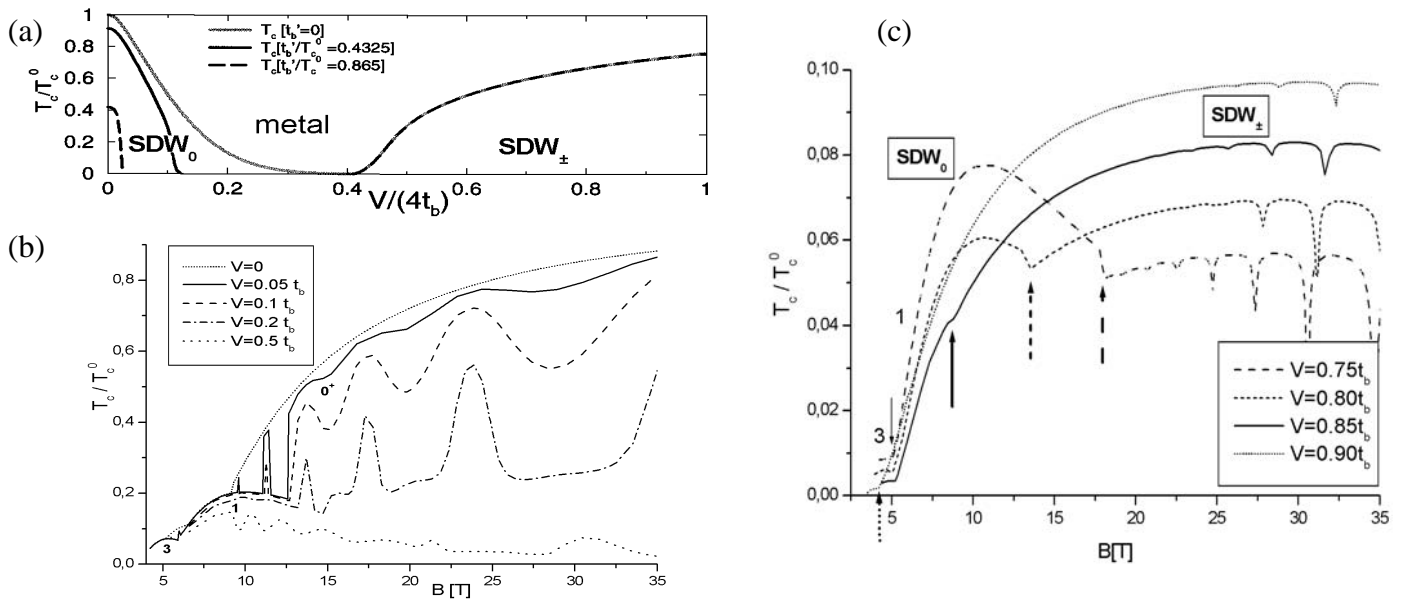

**Figure 1.** Zero magnetic field phase diagram for $t_b$=300K, $T_c^0$=13K and several choices of $t'_b$ (a). Phase diagram in finite magnetic field for $t'_b$=0.03$t_b$ and several choices of $V$ in low-$V$ (b) and intermediate to strong-$V$ limit (c).

**References**
[1] T. Ishiguro, K. Yamaji and G. Saito, Organic Superconductors IIe, (Springer-Verlag, Berlin, 1998).
[2] S. K. McKernan *et al*., Phys. Rev. Lett. **75**, 1630 (1995).
[3] D. LePeleven *et al*., Eur. Phys. J. B**19**, 363 (2001); H.Yoshino *et al*., J. Phys. Soc. Jpn. **68**, 3142 (1999)
[4] J. S. Qualls *et al*., Phys. Rev. B **62**, 12680 (2000); N. Matsunaga *et al*., Phys. Rev. B **62**, 8611 (2000).
[4] T. Osada *et al*., Phys. Rev. Lett. **69**, 1117 (1992).
[5] L. P. Gor'kov and A. G. Lebed, Phys. Rev. B **51**, 3285 (1995); ibid. 1362 (1995);
    A. A. Slutskin and A. M. Kadigrobov, Sov. Phys. Solid State **9**, 138 (1967).
[6] Y. Hasegawa *et al*., J. Phys. Soc. Jpn. **67**, 964 (1998); K. Kishigi, J. Phys. Soc. Jpn. **67**, 3825 (1998).
[7] D. Zanchi and A. Bjelis, Europhys. Lett. **56**, 596 (2001).
[8] K. Sengupta and N. Dupuis, Phys. Rev. B **65**, 035108 (2002).
[9] D. Radić, A. Bjelis and D. Zanchi, cond-mat/0206466; D. Radić *et al*., paper in preparation